\documentclass[final,authoryear,3p,times]{elsarticle}

\usepackage{graphicx}

\usepackage{amssymb}

\usepackage{ulem}

\journal{Phys. Earth Planet. Int.}

\begin{document}

\begin{frontmatter}

\title{Surface atmospheric pressure excitation of the translational mode of the inner core}

\author{S. Rosat\corref{cor1}}
\ead{srosat@unistra.fr}
\author{J.-P. Boy}
\author{Y. Rogister}
\cortext[cor1]{Corresponding author}

\address{Institut de Physique du Globe de Strasbourg; UMR 7516, Universit\'e de Strasbourg/EOST, CNRS - 5 rue Descartes, 67084 Strasbourg, France}

\begin{abstract}
Using hourly atmospheric surface pressure field from ECMWF (European Centre for Medium-Range Weather Forecasts)
and from NCEP (National Centers for Environmental Prediction) Climate Forecast System Reanalysis (CFSR) models, 
we show that atmospheric pressure fluctuations excite the translational oscillation of the inner core, 
the so-called Slichter mode, to the sub-nanogal level at the Earth surface. 
The computation is performed using a normal-mode formalism for a spherical, self-gravitating 
anelastic PREM-like Earth model. We determine the statistical response in the form of power spectral densities of 
the degree-one spherical harmonic components of the observed pressure field. Both hypotheses of inverted and 
non-inverted barometer for the ocean response to pressure forcing are considered. Based on previously computed 
noise levels, we show that the surface excitation amplitude is below the limit of detection of the superconducting gravimeters, making 
the Slichter mode detection a challenging instrumental task for the near future. 
\end{abstract}


\end{frontmatter}


\section{Introduction}
The search for the surface gravity effect of the free translational oscillations of the inner core, the so-called Slichter modes \citep{slichter}, 
has been a subject of observational challenge, particularly since the development of worldwide data from superconducting gravimeters (SGs) of the Global Geodynamics Project \citep{ggp}.
Indeed these relative gravimeters are the most suitable instruments to detect the small signals that would be expected from the Slichter modes \citep{hinderer1995, rosat2003, rosat2004}.

A first claim by \citet{smylie} of a triplet of frequencies that he attributed to the Slichter modes led to a controversy (e.g. \citet{rieutord}). 
This detection has been supported by \citet{courtier} 
and \citet{pagiatakis} but has not been confirmed by other authors \citep{hinderer1995,jensen1995,rosat2006,guo2007,rosatCDW}. 
\citet{crossley92, crossley93} have shown it is necessary to consider dynamic Love numbers to calculate the Slichter mode eigenperiods.
Latest theoretical computation predicts a degenerate (without rotation or ellipticity) eigenperiod of 5.42 h \citep{rogister} for the seismological reference PREM \citep{prem}
 Earth model. A more recent study by \citet{grinfeldwisdom} states that the period could be shorter because of the kinetics of phase transformations at the inner-core boundary (ICB).

The interest raised by the Slichter modes resides in its opportunity to constrain the density jump and the viscosity in the fluid outer core at the ICB. 
The density jump at the ICB is a parameter that constrains the kinetic energy required to power the geodynamo by compositional convection. 
Some discrepancies have been obtained for the value of this parameter. On the one hand, by analyzing seismic PKiKP/PcP phases, 
\citet{koperpyle} found that it should be smaller than 450 kg/m$^3$, later increased to 520 kg/m$^3$ \citep{koperdombro}. On the other hand, using normal modes observation,
 \citet{mastersgubbins} obtained 820 $\pm$ 180 kg/m$^3$. Such differences in the estimate of the ICB density jump have been partially attributed to the uncertainties associated 
with the seismic noise \citep{tkalcic}. A model that satisfies both the constraints set by powering the geodynamo 
with a reasonable heat flux from the core, and PKP traveltimes and normal mode frequencies has been proposed by \citet{gubbins} with a large overall density jump between 
the inner and outer cores of 800 kg/m$^3$ and a sharp density jump of 600 kg/m$^3$ at the ICB itself. In the following we will adopt the PREM value of 600 kg/m$^3$.

The non-detection of the Slichter modes raises the question of their expected amplitude, their damping and the possible mechanisms to excite them. 
A certain number of papers have considered the damping of the inner core oscillation through anelasticity of the inner core and mantle \citep{crossley91}, 
through viscous dissipation in the outer core \citep{mathewsguo} or through magnetic dissipation \citep{buffett}. 
\citet{guo2007} and \citet{rosatCDW} have summarized the theoretical Q values expected for the Slichter mode. 
\citet{greff} have concluded that it should most probably be equal to or larger than 2000. Various sources of excitation have been previously considered. 
The seismic excitation has been studied by \citet{smith}, \citet{crossley} and \citet{rosat2007}. 
They have shown that earthquakes cannot excite the Slichter modes to a level sufficient for the SGs to detect the induced surface gravity effect. 
For instance, even for the 1960 $M_w=9.6$ Chilean event the induced surface gravity effect does not reach the nanogal level (1 nGal$=10^{-2}$ nm/s$^2$). 
Surficial pressure flow acting at the ICB and generated within the fluid outer core has been considered by \citet{greff} and \citet{rosatrogister} as a possible excitation mechanism. 
However, the flow in the core at a timescale of a few hours is too poorly constrained to provide reliable predictions of the amplitude of the Slichter modes.
\citet{rosatrogister} have investigated the excitation of the Slichter modes by the impact of a meteoroid, which they treated as a surficial seismic source. 
For the biggest known past collision associated to the Chicxulub crater in Mexico with a corresponding moment-magnitude $M_w=9.6$, 
the surface excitation amplitude of the Slichter mode was barely 0.0067 nm/s$^2$ $=$ 0.67 nGal. 
Nowadays, a similar collision would therefore not excite the Slichter modes to a detectable level.
The degree-one surface load has also been investigated by \citet{rosatrogister}. They showed that a Gaussian-type zonal degree-one pressure flow of 4.5 hPa applied during 1.5 hour
 would excite the Slichter mode and induce a surface gravity perturbation of 2 nGal which should be detectable by SGs \citep{rosathinderer}. 
This determination was based on a purely analytical model of surface pressure.

In this paper we will use hourly surface pressure data provided by two different meteorological centers and show that the surface atmospheric pressure fluctuations 
can only excite the Slichter modes to an amplitude below the limit of detection of current SGs. 

\section{Excitation by a continuous surface load}

\arraycolsep1.5pt
In this Section, we consider a spherical Earth model, for which the frequencies of the three Slichter modes degenerate into a single frequency, 
and establish a formula for the spectral energy of the amplitude of the mode when it is excited by a surface load.

Developed in a surface spherical harmonics expansion, a degree-one load $\sigma_0$ contains three terms:
\begin{equation}
\sigma_0(t;\theta, \phi) =\sigma_{10}(t) \cos\theta + (\sigma_{11}^c(t) \cos\phi + {\sigma}_{11}^s(t) \sin\phi) \sin\theta ,
\label{load}
\end{equation}
where $\theta$ and $\phi$ are the colatitude and longitude, respectively. The Green function formalism suited for surface-load problems 
\citep{farrell} has been generalized to the visco-elastic case by \citet{tromp} and has been established for the degree-one Slichter mode by \citet{rosatrogister}. 

The degree-one radial displacement due to load (\ref{load}) is given by
\begin{eqnarray}
& & u_r(r,\theta,\phi;t)  =  \frac{r_s^2 U(r)}{i\nu} [U(r_s)g_0+P(r_s)] \nonumber \\ 
& & \lbrack \int_{-\infty}^{t}e^{i\nu t'} (\sigma_{10}(t') \cos\theta + \sigma_{11}^c(t') \sin\theta \cos\phi 
 + { \sigma}_{11}^s(t') \sin\theta \sin\phi) dt' \rbrack ,
 \label{RadialDisplacement(t)}
\end{eqnarray} 
and the perturbation of the surface gravity is
\begin{eqnarray}
\label{deltag}
& & \Delta g(r,\theta,\phi;t)= \frac{r_s^2 U(r)}{i\nu} [U(r_s)g_0+P(r_s)] \nonumber \\  
& & \lbrack \int_{-\infty}^{t}e^{i\nu t'} (\sigma_{10}(t') \cos\theta + \sigma_{11}^c(t') \sin\theta \cos\phi 
 + { \sigma}_{11}^s(t') \sin\theta \sin\phi) dt' \rbrack  \nonumber \\  
& & [-\omega^2U(r_s)+\frac{2}{r_s}g_0 U(r_s)+\frac{2}{r_s}P(r_s)].
\end{eqnarray}
In the last two equations, $U$ and $P$ are, respectively, the radial displacement and perturbation of the 
gravity potential associated to the Slichter mode, $r_s$ is the Earth radius, 
and $\nu=\omega_0(1+i/2Q)$ is the complex frequency. As in \citet{rosatrogister}, we adopt a quality factor $Q = 2000$ 
and a period $T = {2\pi}/{\omega_0} = 5.42$ h for a PREM-like Earth's model.

The sources of excitation we consider are continuous pressure variations at the surface. 
A similar problem was treated by \citet{tanimotoum} and \citet{fukao} for the atmospheric excitation of normal modes where the sources were considered as stochastic quantities in space and time. 
As we use a harmonic spherical decomposition of the pressure field, the correlation in space depends on the harmonic degree, here the degree-one component of wavelength $2 \pi r_s$. 
The correlation in time is performed in the spectral domain.

As a consequence we introduce the energy spectrum of the degree-one pressure fluctuations
\begin{eqnarray}
\label{sp}
S_p(\theta,\phi;\omega) = {\hat \sigma_0}(\omega; \theta,\phi) {\hat \sigma_0^*}(\omega; \theta,\phi),
\end{eqnarray}
and the energy spectrum of the radial displacement
\begin{eqnarray}
S(r,\theta,\phi;\omega) = {\hat u}_r(r,\theta,\phi;\omega) {\hat u}_r^*(r,\theta,\phi;\omega), \nonumber 
\end{eqnarray}
where ${\hat u}_r(r,\theta,\phi;\omega)$ is the Fourier transform of $u_r(r,\theta,\phi;t)$, 
${\hat \sigma_0}(\omega; \theta,\phi)$ is the Fourier transform of $\sigma_0(t; \theta,\phi)$ and $^*$ denotes the complex conjugate.
The Fourier transform of Eq. (\ref{RadialDisplacement(t)}) is 
\begin{eqnarray}
{\hat u}_r(r,\theta,\phi;\omega)&=& \frac{r_s^2 U(r)}{i\nu} [U(r_s)g_0+P(r_s)] 
\int_{-\infty}^{+\infty} e^{-i\omega t} \lbrack \int_{-\infty}^{t}e^{i\nu_kt'} \nonumber \\
& & (\sigma_{10}(t') \cos\theta + \sigma_{11}^c(t') \sin\theta \cos\phi 
 + { \sigma}_{11}^s(t') \sin\theta \sin\phi) dt' \rbrack dt \nonumber \\
&=& \frac{r_s^2 U(r)}{\omega_0(1+\frac{1}{4Q^2})} [U(r_s)g_0+P(r_s)] 
 \frac{\omega_0(1-\frac{1}{4Q^2})-\omega+\frac{i}{2Q}(\omega-2\omega_0)}{\frac{\omega_0^2}{4Q^2}+(\omega_0-\omega)^2} \nonumber \\ 
& & ({\hat \sigma_{10}}(\omega)\cos\theta + {\hat \sigma_{11}^c}(\omega)\sin\theta \cos\phi + {\hat \sigma_{11}^s}(\omega)\sin\theta \sin\phi), \nonumber
\end{eqnarray}
and, therefore, we have
\begin{eqnarray}
\label{exdispl}
S(r,\theta,\phi;\omega)&=& \frac{r_s^4 U^2(r)}{\omega_0^2(1+\frac{1}{4Q^2})^2} [U(r_s)g_0+P(r_s)]^2 \nonumber \\ 
& & \frac{\lbrack \omega_0(1-\frac{1}{4Q^2})-\omega \rbrack^2+\frac{(\omega-2\omega_0)^2}{4Q^2}} 
{\lbrack \frac{\omega_0^2}{4Q^2}+(\omega_0-\omega)^2\rbrack ^2} S_p(\theta,\phi;\omega). 
\end{eqnarray}
From Eq.(\ref{exdispl}) and (\ref{deltag}), we obtain the spectral energy of gravity $S_g$ for the excitation of the Slichter mode by a surface fluid layer
\begin{eqnarray}
\label{exg}
S_g(r,\theta,\phi;\omega)&=& \frac{r_s^4 (-\omega^2U(r)+\frac{2}{r}g_0U(r)+\frac{2}{r}P(r))^2}{\omega_0^2(1+\frac{1}{4Q^2})^2} [U(r_s)g_0+P(r_s)]^2 \nonumber \\ 
& & \frac{\lbrack \omega_0(1-\frac{1}{4Q^2})-\omega \rbrack^2+\frac{(\omega-2\omega_0)^2}{4Q^2}} 
{\lbrack \frac{\omega_0^2}{4Q^2}+(\omega_0-\omega)^2\rbrack ^2} S_p(\theta,\phi;\omega).
\end{eqnarray}

\section{Atmospheric pressure variations}
\label{pressuredata}

The operational model of the European Centre for Medium-Range Weather Forecasts (ECMWF) is usually available at 3-hourly temporal resolution,
the spatial resolution varying from about 35 km in 2002 to 12 km since 2009. This is clearly not sufficient to investigate the Slichter mode excitation. 
However, during the period of CONT08 measurements campaign (August $12-26$, 2008), atmospheric analysis data were provided by the ECMWF also on an hourly basis. 
CONT08 provided 2 weeks of continuous Very Long Baseline Interferometry (VLBI) observations for the study, 
among other goals, of daily and sub-daily variations in Earth rotation \citep{cont08}. 
We take advantage of this higher-than-usual temporal resolution to compute the excitation of the Slichter mode by the surface pressure fluctuations.
To do so, we first extract the degree-one coefficients of the surface pressure during this period, considering both an inverted and a non-inverted barometer 
response of the oceans to air pressure variations \citep{wunsch}. Both hypotheses have the advantage to give simple responses of the oceans to atmospheric forcing, 
even if at such high frequencies, static responses are known to be inadequate. The use of a dynamic response of the ocean \citep{dynoc} would lead to 
more accurate results but we would need a forcing (pressure and winds) of the oceans at hourly time scales which is not available.
 
The degree-1 surface pressure changes contain three terms: 
\begin{eqnarray}
P_0 = C_{10}(t)\cos\theta + (C_{11}(t) \cos\phi + S_{11}(t) \sin\phi) \sin\theta. \nonumber
\end{eqnarray}
From it, we can estimate the surface mass density by $\sigma_{0} =P_{0}/g_0$ where $g_0$ is the mean surface gravity and compute the energy spectrum $S_p$ as defined in Eq.(\ref{sp}).

\begin{figure}[!ht]
\centering
\includegraphics[width=6cm]{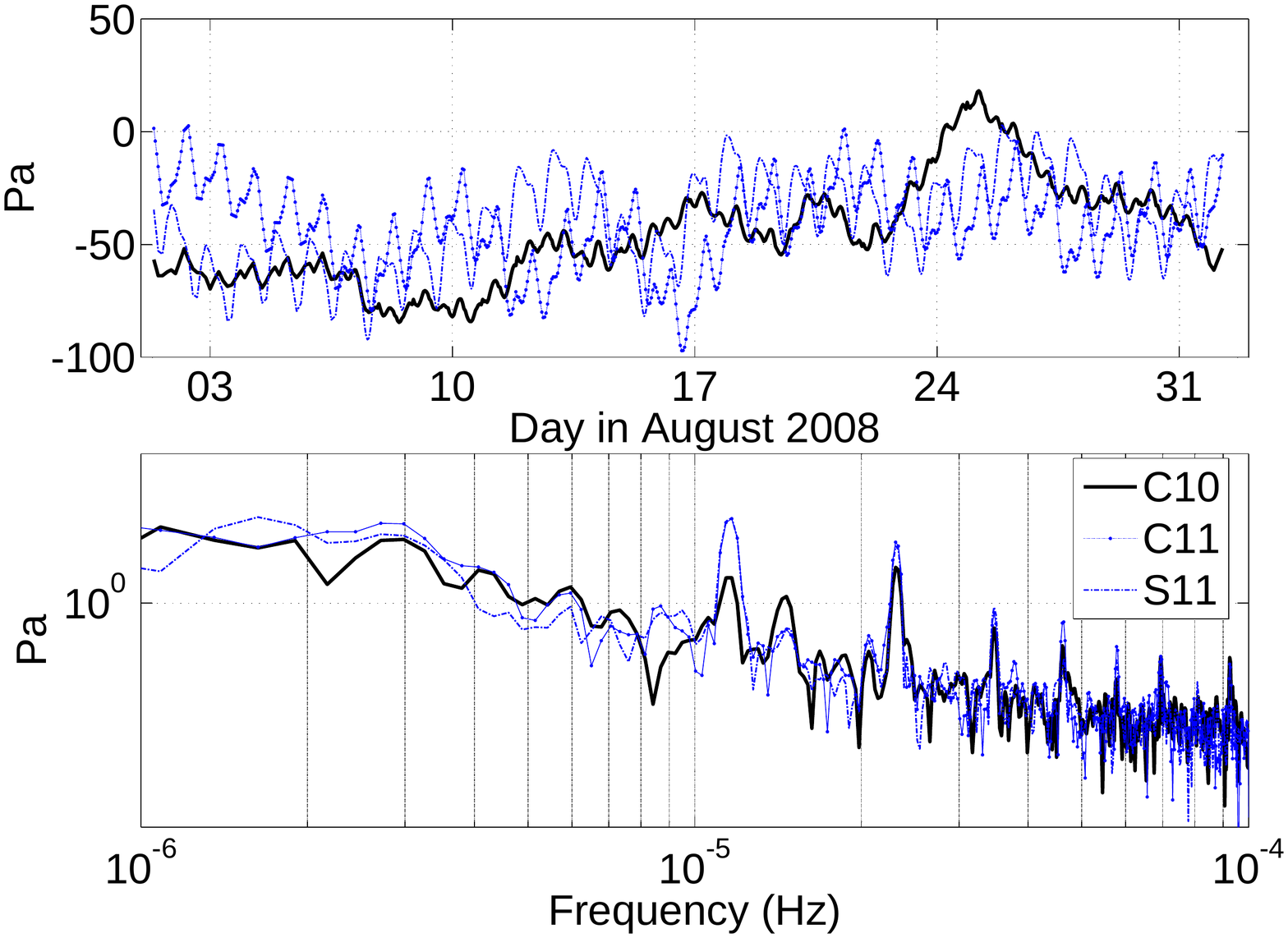}
\includegraphics[width=6cm]{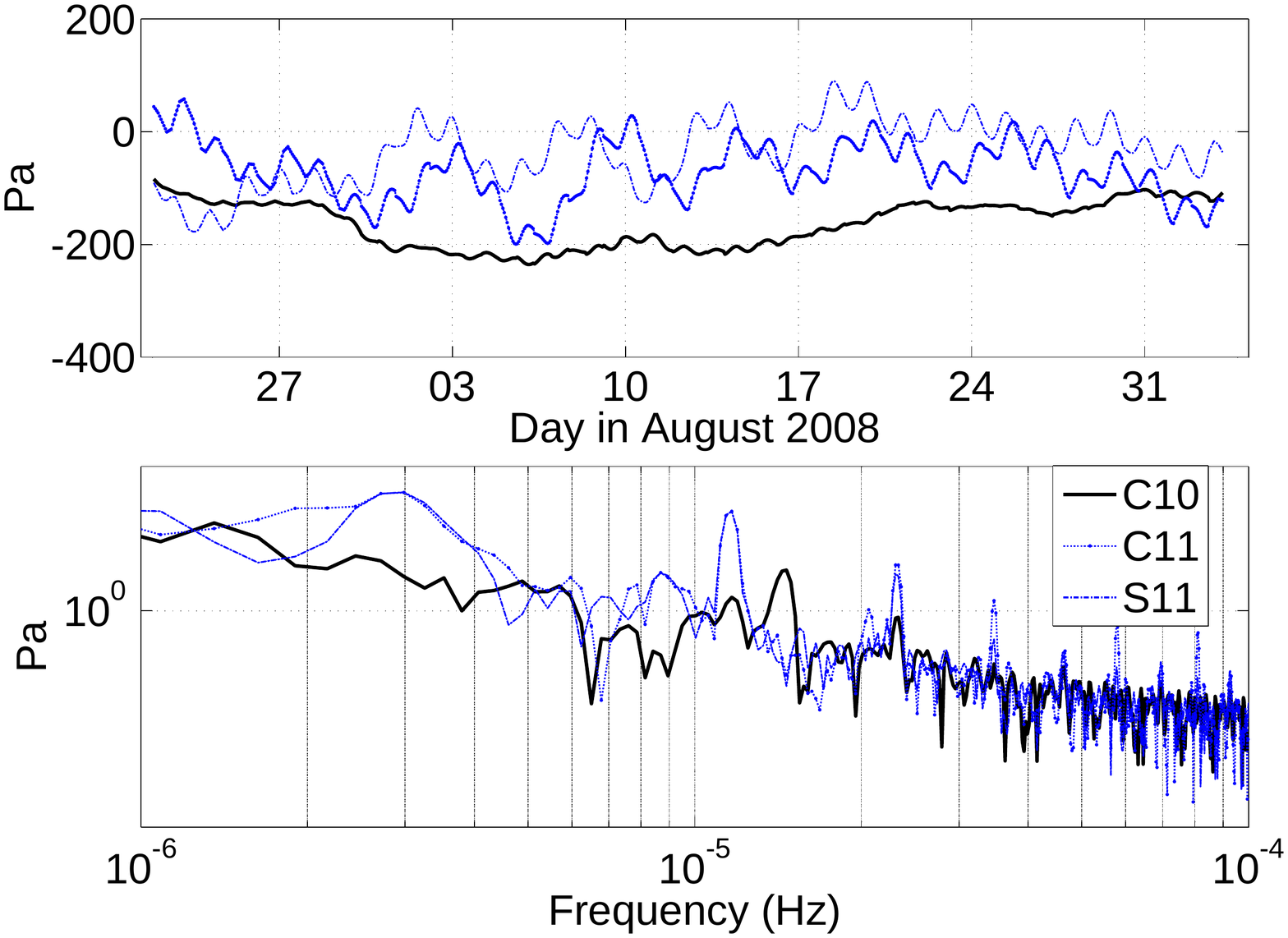}
\caption{Time-variations (upper) and amplitude spectra (lower) of the harmonic degree-one 
coefficients of the ECMWF atmospheric pressure field during August 2008 for an inverted ({\it left}) and a non-inverted ({\it right}) barometer response of the oceans.}
\label{fig:time_fft_IB}
\end{figure}

\begin{figure}[!ht]
\centering
\noindent\includegraphics[width=10cm]{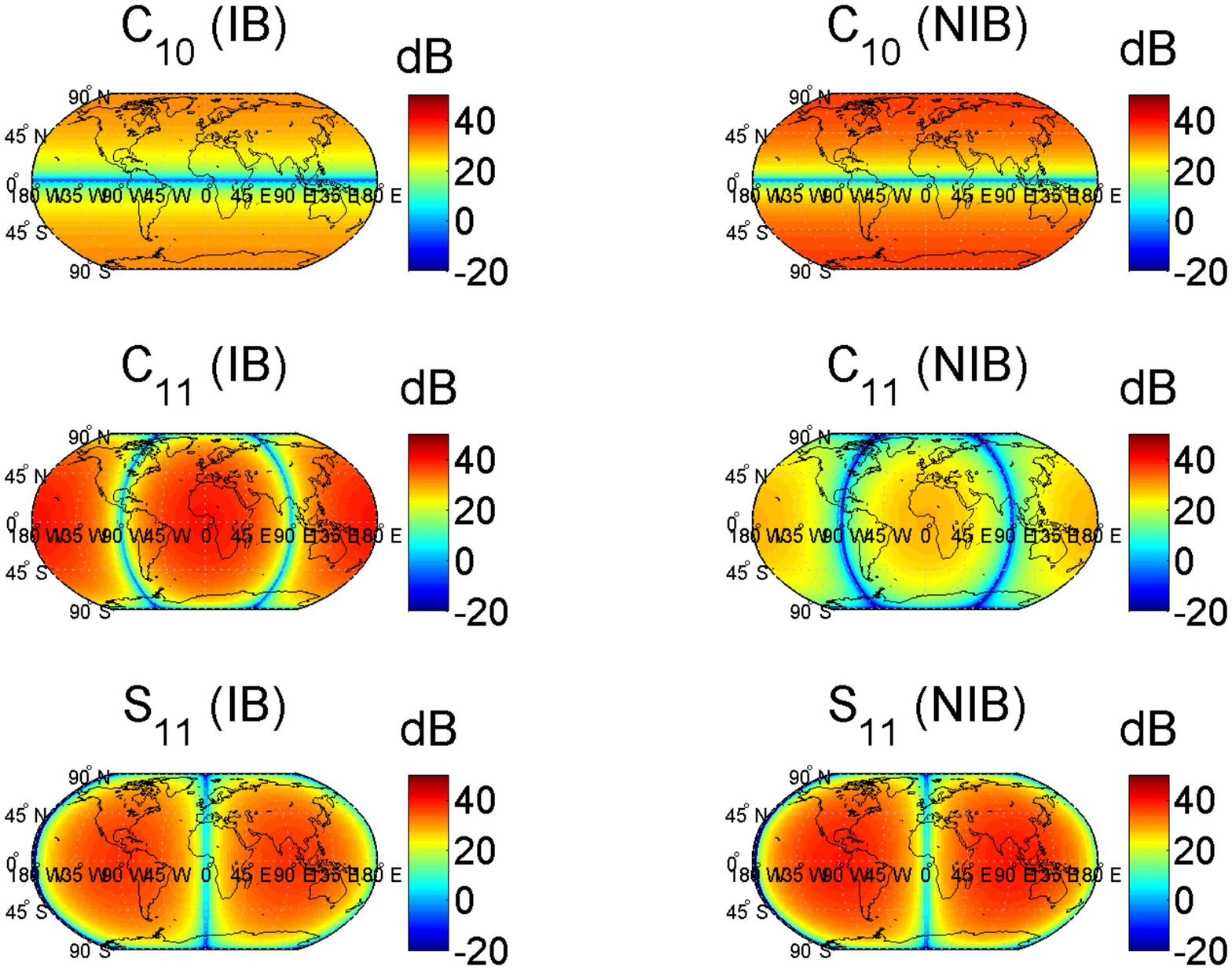}
\caption{Power spectral densities of the harmonic degree-one coefficients of the ECMWF atmospheric pressure field 
during August 2008 for an inverted and a non-inverted barometer response of the oceans. Unit is decibel relatively to $Pa^2/Hz$.}
\label{fig:mapsPSD}
\end{figure}

The time-variations and Fourier amplitude spectra of the harmonic degree-one coefficients $C_{10}$, $C_{11}$ and $S_{11}$ 
are plotted in Fig.\ref{fig:time_fft_IB} for the IB and non-IB hypotheses.
The power spectral densities (PSDs) computed over the whole period considered here (August 2008) are represented in Fig.\ref{fig:mapsPSD} for both oceanic responses.

\begin{figure}[!ht]
\centering
\noindent\includegraphics[width=10cm]{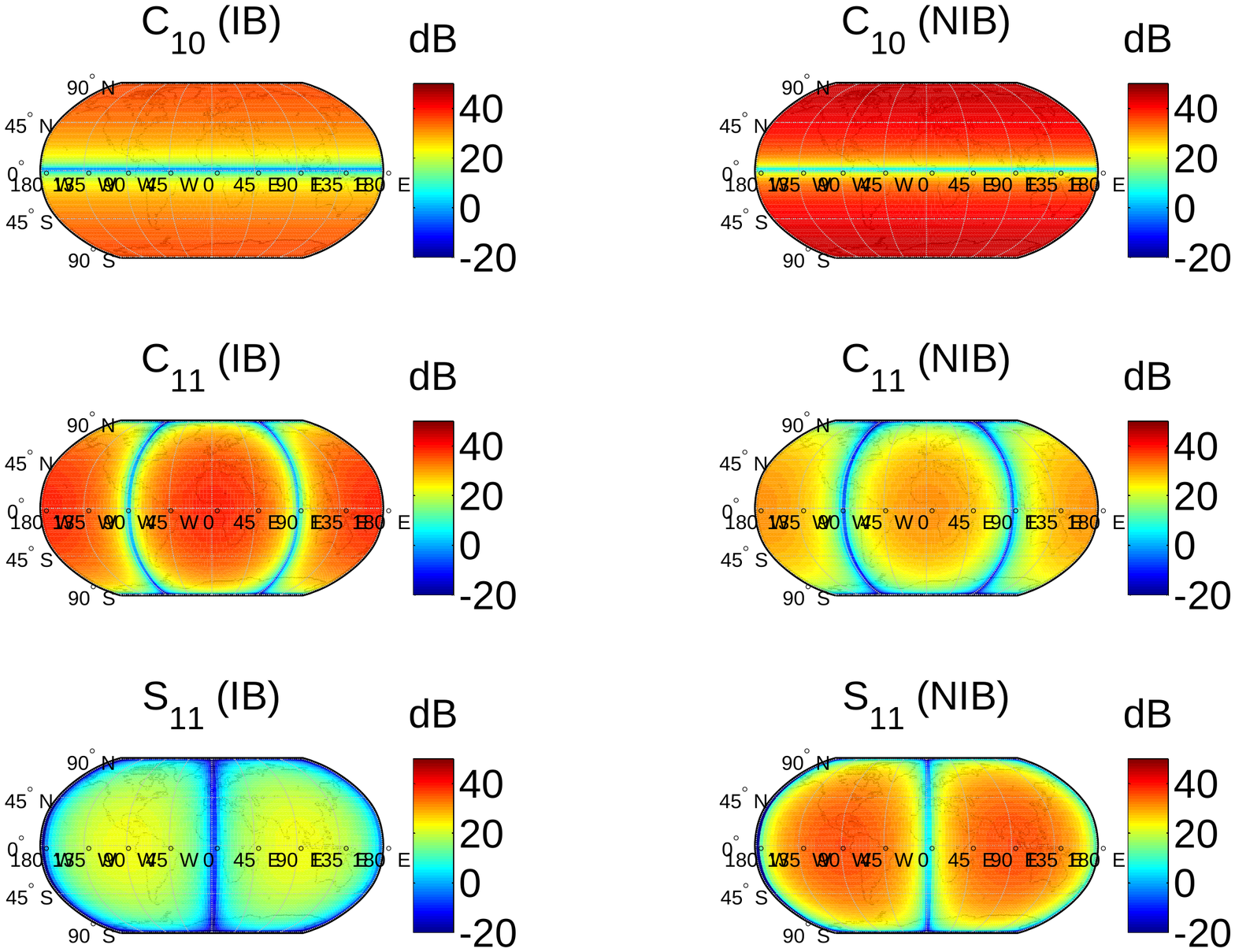}
\caption{Power spectral densities of the harmonic degree-one coefficients of the NCEP/CFSR atmospheric pressure field 
during August 2008 for an inverted and a non-inverted barometer response of the oceans. Unit is decibel relatively to $Pa^2/Hz$.}
\label{fig:mapsPSDncep}
\end{figure}

We also consider the NCEP (National Centers for Environmental Prediction) Climate Forecast System Reanalysis (CFSR) model \citep{ncep}, for which hourly
surface pressure is available with a spatial resolution of about 0.3$^o$ before 2010, and 0.2$^o$ after. NCEP/CFSR and ECMWF models used different assimilation schemes,
e.g. 4D variational analysis for ECMWF, and a 3D variational analysis for NCEP/CFSR every 6~hours. The temporal continuity of pressure is therefore enforced in the ECMWF model,
whereas 6-hourly assimilation steps can sometimes be seen in the NCEP/CFSR pressure series, for certain areas and time.
The power spectral densities of the degree-one NCEP/CFSR surface pressure field computed over August 2008 are represented in Fig. \ref{fig:mapsPSDncep} for both oceanic responses.

\section{Excitation amplitude of the Slichter mode}

Using equations (\ref{deltag}) and (\ref{exg}) we can compute the surface gravity perturbation induced by the Slichter mode excited 
by the degree-one ECMWF and NCEP/CFSR surface atmospheric pressure variations during August 2008. 
The computation is performed with the eigenfunctions obtained for a spherical, self-gravitating anelastic PREM-like Earth model as in \citet{rogister}. 
We remove the 3 km-thick global ocean from the PREM model because the ocean response is already included in the degree-one atmospheric coefficients.

\begin{figure}[!ht]
\centering
\includegraphics[width=10cm]{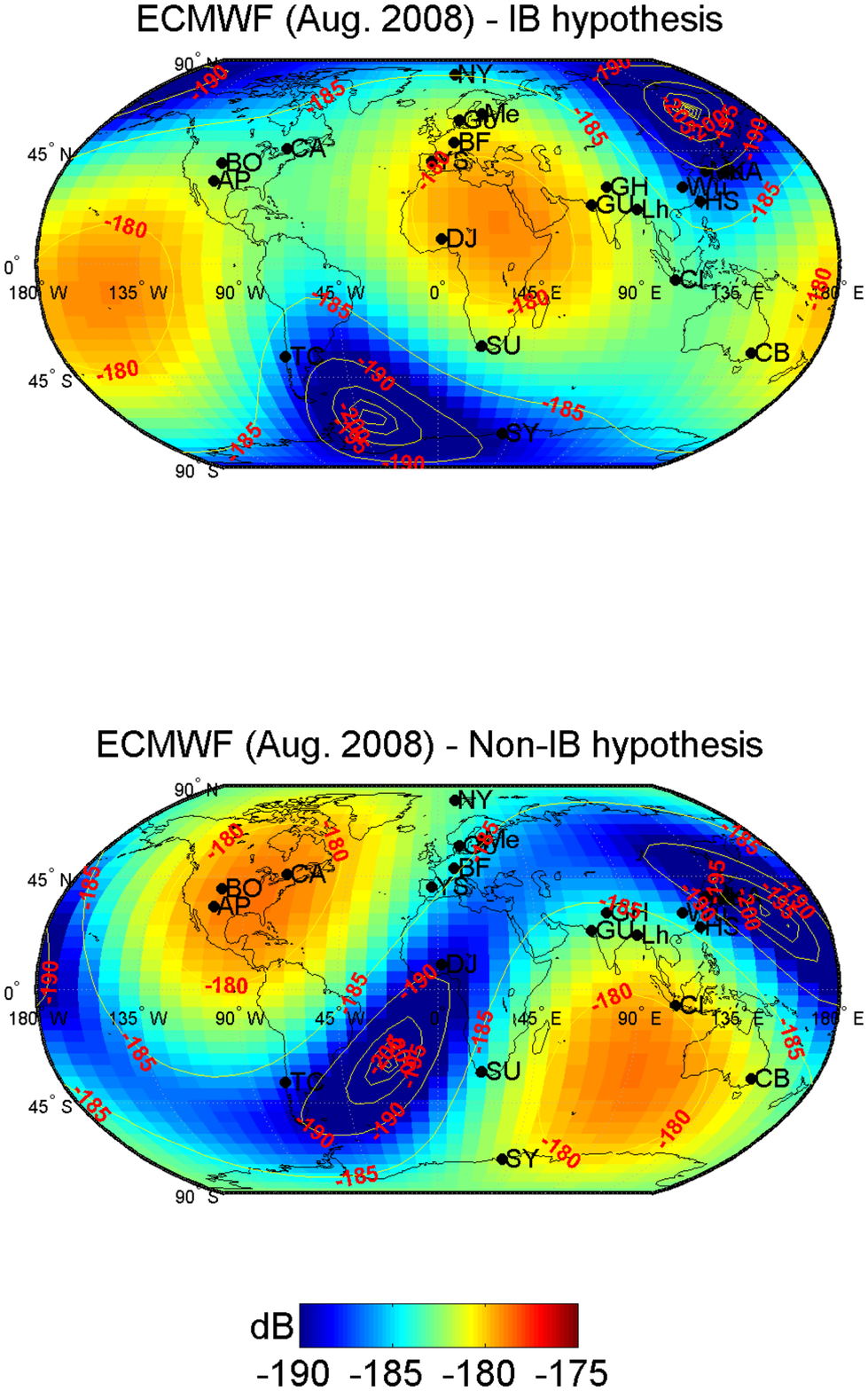}
\caption{Power spectral density of the surface gravity effect induced by the Slichter mode excited by the degree-one coefficients of the ECMWF pressure field 
during August 2008 for an inverted ({\it left}) and a non-inverted ({\it right}) barometer response of the oceans. 
Unit is decibel relatively to $(m/s^2)^2/Hz$. Some superconducting gravimeter sites are indicated.}
\label{fig:dgIB-NIB_maps_ECMWF}
\end{figure}

The power spectral densities of the surface gravity effect induced by the Slichter mode excited by the ECMWF atmospheric data are plotted in
Figs \ref{fig:dgIB-NIB_maps_ECMWF} for an inverted and a non-inverted barometer response of the oceans. 
The PSD is given in decibels to enable an easy comparison with previous SG noise level studies (e.g. \citet{rosathinderer}). 
NCEP/CFSR weather solutions give a similar surface excitation amplitude less than -175 dB.

\begin{figure}[!ht]
\centering
\includegraphics[width=10cm]{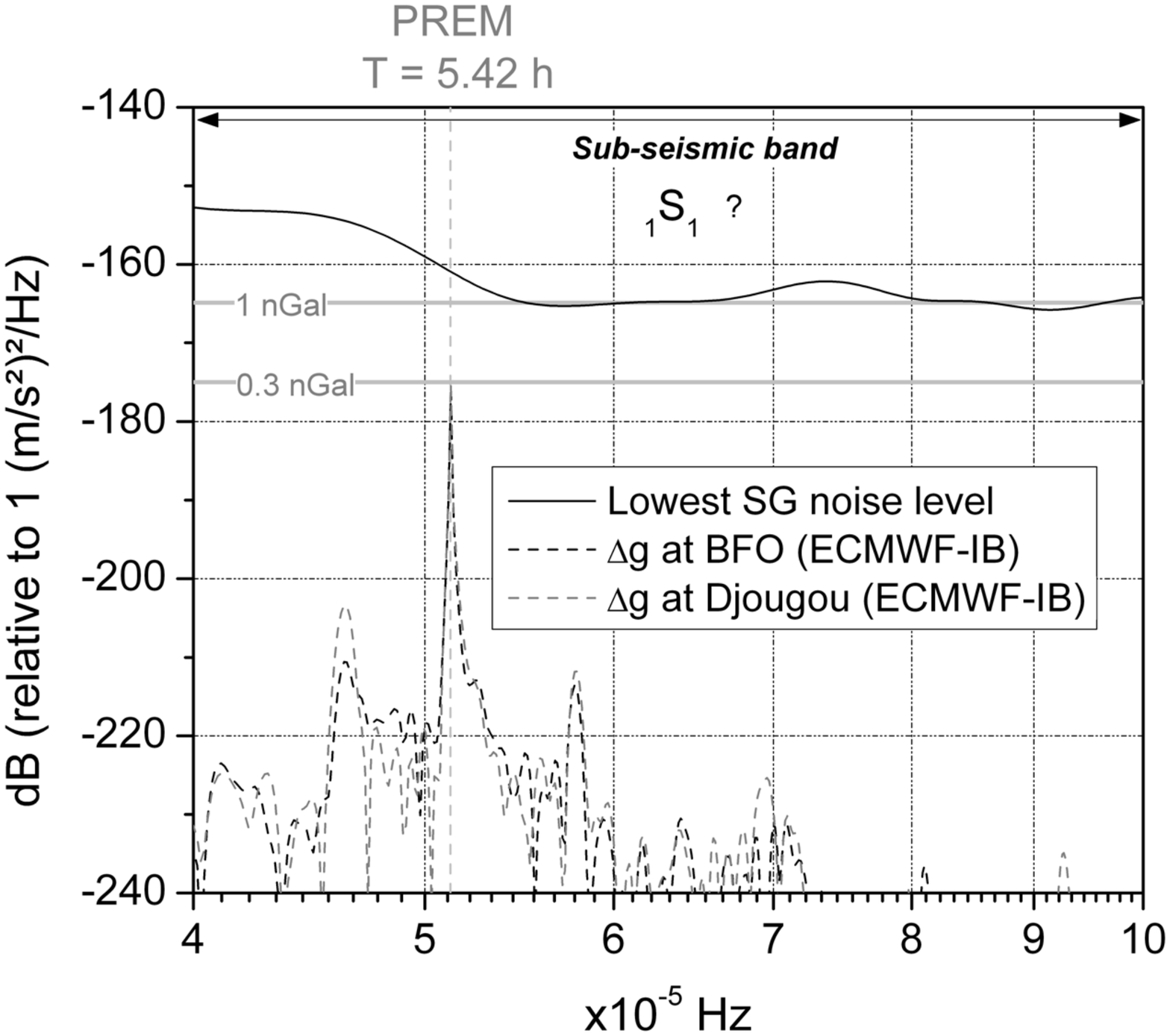}
\includegraphics[width=10cm]{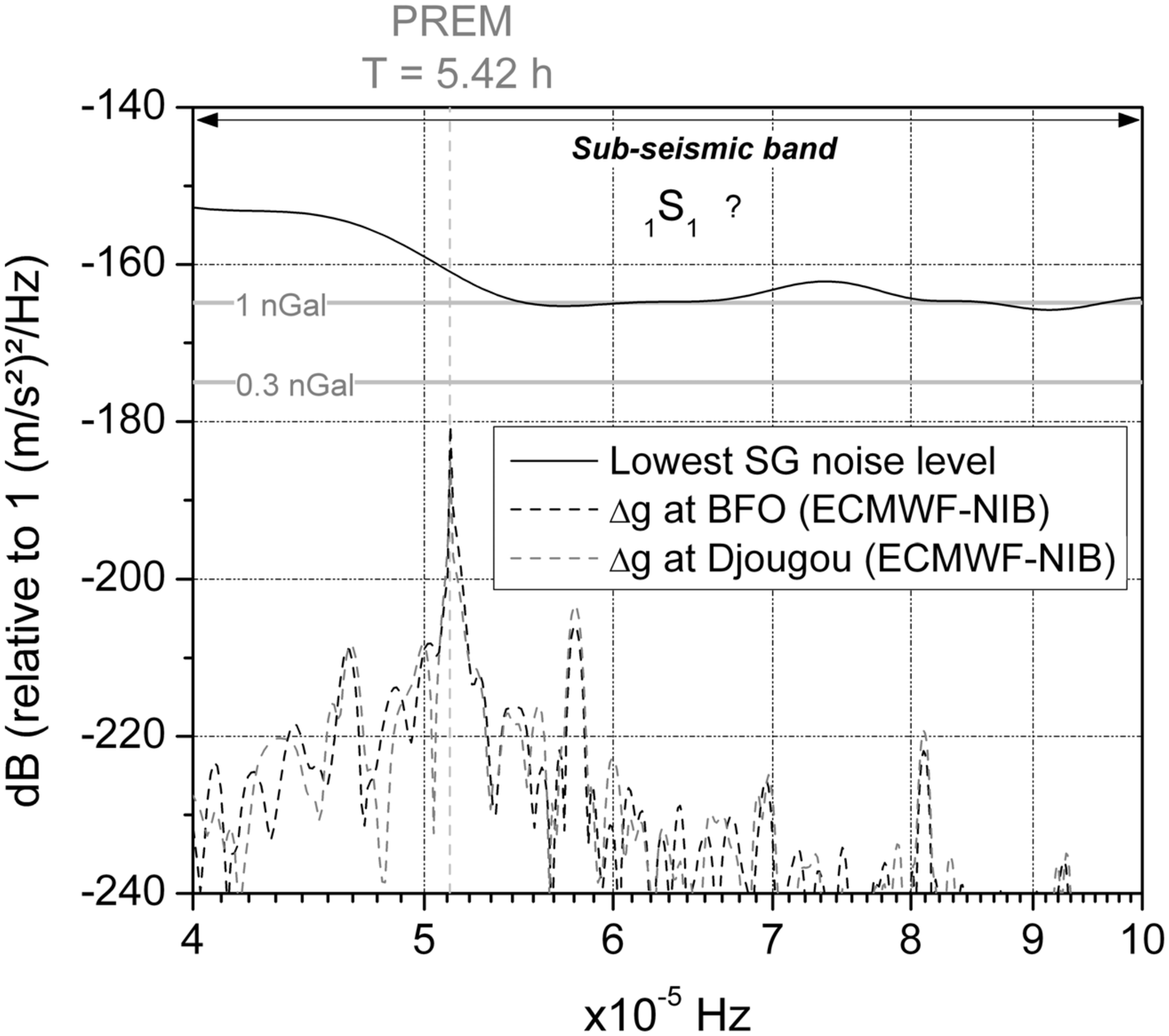}
\caption{Power spectral densities of the surface gravity effect at Djougou and BFO induced by the Slichter mode excited by the degree-one coefficients of the ECMWF atmosphere 
during August 2008 for an inverted ({\it top}) and non-inverted ({\it bottom}) barometer response of the oceans. Unit is decibel relatively to $(m/s^2)^2/Hz$. 
The best SG noise level and the levels corresponding to the 1 nGal and 0.3 nGal signals are indicated.}
\label{fig:dg_ECMWF_B1DJ_nGal}
\end{figure}

We also consider the PSDs of the excitation amplitude at the SG sites Djougou (Benin) and BFO (Black Forest Observatory, Germany) 
for both oceanic responses in Fig. \ref{fig:dg_ECMWF_B1DJ_nGal}. 
According to Fig. \ref{fig:dgIB-NIB_maps_ECMWF} the Djougou site turns out to be located at a maximum of excitation amplitude in the case of an inverted barometer response of the oceans. 
BFO is the SG site with the lowest noise level at sub-seismic frequencies \citep{rosathinderer}. The later noise level is also plotted in Fig. \ref{fig:dg_ECMWF_B1DJ_nGal}. 
In that figure, we can see that the excitation amplitude at Djougou reaches -175 dB. 
For an undamped harmonic signal of amplitude $A$ the PSD is defined by $A^2NT_0/4$ where $N$ is the number of samples and $T_0$ the sampling interval. 
Consequently, assuming a 15-day time duration with a sampling rate of 1 min, a PSD amplitude of -175 dB
 corresponds to a harmonic signal of 0.3 nGal, which is clearly below the 1 nGal detection threshold and below the best SG noise level. 
A decrease of noise by a factor 3 would be necessary to be able to detect such a sub-nanogal effect. 
Stacking $n$ worldwide SGs of low noise levels would improve the signal-to-noise ratio by a factor $\sqrt n$. 
Supposing that we had a large number of SG sites with equal noise levels (same as BFO SG noise level), 
then we would need to stack 10 datasets to improve the SNR by 10 dB, so as to reach the nanogal level.

\begin{figure}[!ht]
\centering
\noindent\includegraphics[width=16cm]{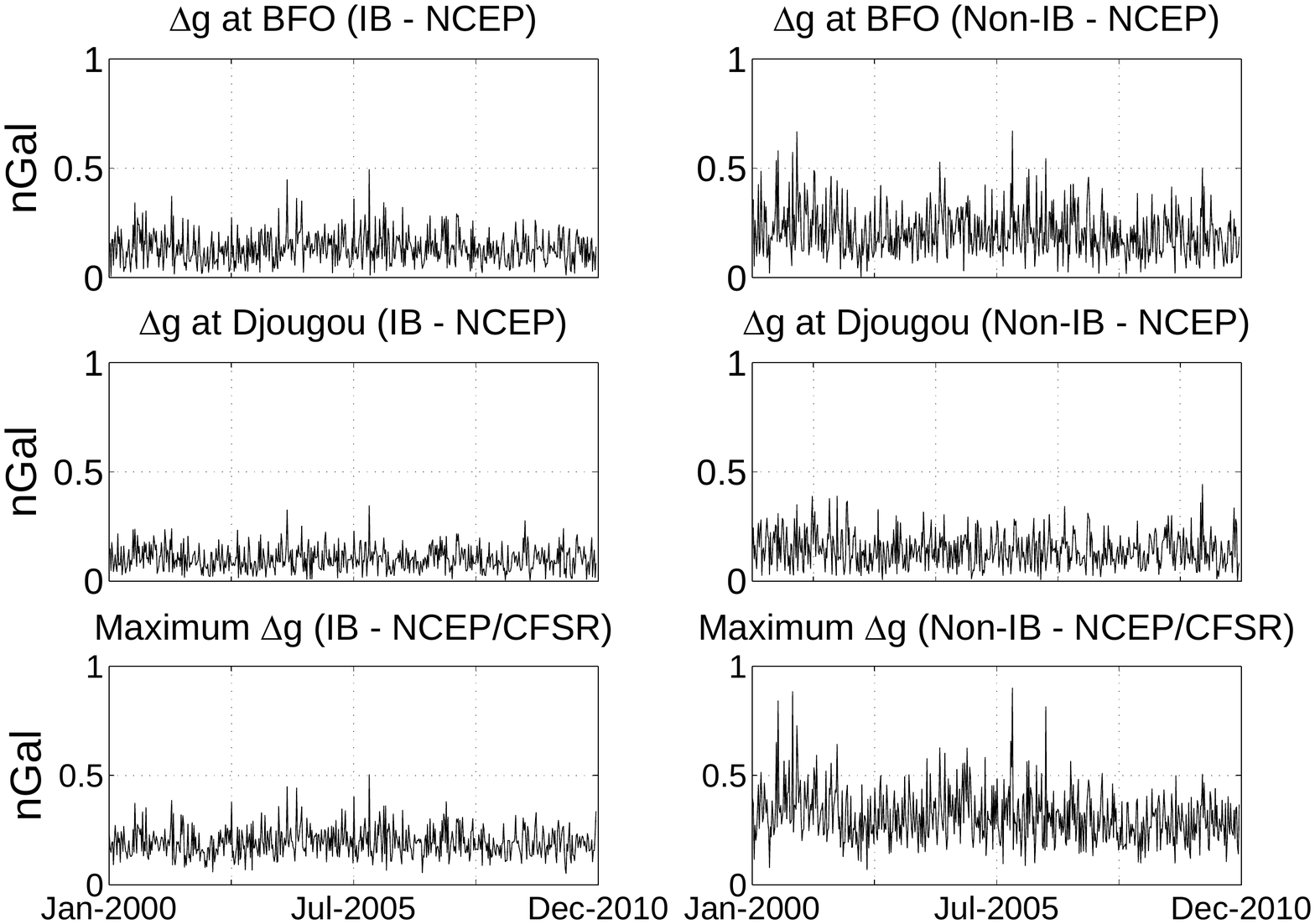}
\caption{Time-varying surface gravity effects induced by the Slichter mode excited by the degree-one coefficients of the NCEP/CFSR atmospheric model 
from 2000 until 2011 for an inverted and a non-inverted barometer response of the oceans at BFO (top), at Djougou (middle) and maximum excitation amplitudes (bottom).}
\label{fig:dg_BFO-DJ_NCEP_11yr}
\end{figure}

Both ECMWF and NCEP atmospheric pressure data lead to a similar, presently undetectable, excitation amplitude for the Slichter mode during August 2008. 
However, as we have at our disposal 11 years of hourly NCEP/CFSR surface pressure field, we can look at the time-variations of the excitation amplitude of the Slichter mode over the full period.
NCEP atmospheric pressure data are assimilated every 6 h introducing an artificial periodic signal. 
In order to avoid the contamination by this 6 h-oscillation in spectral domain on the Slichter mode period of 5.42 h, we need a data length of 2.5 days at least. 
We consider time-windows of 15 days shifted by 7 days and compute the surface excitation amplitude of the Slichter mode at the BFO and Djougou superconducting gravimeter sites and 
at location on Earth for which the excitation amplitude is maximum (Fig.\ref{fig:dg_BFO-DJ_NCEP_11yr}). Note that this location of maximum amplitude is also varying in time. 
We can see that the excitation amplitude is larger than 0.4 nGal at BFO for instance between January and March 2004 
and in November 2005 for both oceanic responses. There is also a peak of excitation at Djougou in November 2005 and between January and March 2004 but only for an IB-hypothesis.
However, during these 11 years between 2000 and 2011, the maximum surface excitation amplitude stays below 0.7 nGal.

As a consequence we can conclude that the degree-one surface pressure variations are a possible source of excitation for the Slichter mode but the induced surface gravity effect is too 
weak to be detected by current SGs. 

\section{Conclusion}
Using a normal mode formalism, we have computed the surface gravity perturbations induced by a continuous excitation of the Slichter mode 
by atmospheric degree-one pressure variations provided by two meteorological centers: ECMWF and NCEP/CFSR. 
Both inverted and non-inverted barometer responses of the oceans to the atmospheric load have been employed. 
We have shown that the induced surface gravity signal does not reach the nanogal level, which is considered as being the level of detection of present SGs. 

The surficial degree-one pressure variations are a probable source of excitation of the Slichter mode but the weak induced surface amplitude 
is one additional reason why this translational mode core has never been detected. An instrumental challenge for the future gravimeters would be to further decrease their noise levels.

Another source of possible excitation that has not been investigated yet is the dynamic response of the oceans. The oceans are known to be a source of continuous excitation of the fundamental 
seismic modes \citep{webb2007, tanimoto2007, webb2008}. So a further study would require to improve the response of the oceans.

\section{Acknowledgments}
We would like to thank two anonymous reviewers for their comments on this work. We acknowledge the use of meteorological data of the ECMWF and NCEP.


\end{document}